\documentclass[reviewcopy]{elsarticle}

\usepackage[reviewcopy]{adndt}
\usepackage{longtable}

\usepackage{amsmath}

\usepackage{amssymb}

\biboptions{square,sort&compress}
\bibpunct[]{[}{]}{,}{n}{}{;}
\begin{document}

\begin{frontmatter}

\journal{Atomic Data and Nuclear Data Tables}

%% Author, fill in article title here

\title{Radiative rates for  E1, E2, M1, and M2 transitions in Ne-like Cu XX, Zn XXI and Ga XXII}%Template for \normalfont\textsc{Atomic Data and Nuclear Data Tables}}

%% Fill in author list here
  \author[One]{Kanti M. Aggarwal\fnref{}\corref{cor1}}
  \ead{K.Aggarwal@qub.ac.uk}

%  \author[One]{Francis P. Keenan}%\fnref{}}

 % \author[Two]{C. Author} 

  \cortext[cor1]{Corresponding author.}
%  \fntext[X]{{\em e-mail address}: K.Aggarwal@qub.ac.uk}
%  \fntext[Y]{{\em e-mail address}: F.Keenan@qub.ac.uk}

  \address[One]{Astrophysics Research Centre, School of Mathematics and Physics, Queen's University Belfast,\\Belfast BT7 1NN,
Northern Ireland, UK}

%  \address[Two]{Second Address First Line\\
%    Second Address Second Line}

\date{16.12.2002} %please do not use \today, use actual date of version

\begin{abstract}  
Energy levels, radiative rates and lifetimes are reported for the lowest 139 levels of three Ne-like ions, namely Cu~XX, Zn~XXI and Ga~XXII. These levels mostly belong to the  2s$^2$2p$^6$,  2s$^2$2p$^5$3$\ell$, 2s2p$^6$3$\ell$, 2s$^2$2p$^5$4$\ell$, 2s2p$^6$4$\ell$, and 2s$^2$2p$^5$5$\ell$ configurations. For the calculations  the general-purpose relativistic atomic structure package (GRASP) has been adopted. Comparisons are made with  earlier available theoretical and experimental results, particularly among the lowest 27 levels of the 2s$^2$2p$^6$ and  2s$^2$2p$^5$3$\ell$  configurations. Due to paucity of similar data for higher lying levels, analogous   calculations have also been performed with the flexible atomic code (FAC). These  calculations help in assessing the accuracy of our calculated results, especially for the energy levels.  \\ \\

{\em Received}: 15 February 2018, {\em Accepted}: 9 March 2018

\vspace{0.5 cm}
\end{abstract}

\end{frontmatter}

%%% Keywords and subject classification are not used in ADNDT 
%%%\begin{keywords}
%%%Insert list of keywords here.
%%%\end{keywords}

%%% The table of contents should start a new page. This command will
%%% automatically pull all the unstarred \section, \subsection and
%%% \subsubsection titles into the Contents. Starred versions need to be
%%% done manually using
%%%            \addcontentsline{toc}{[[sub]sub]section}{Section title}
%%% at the correct place. Examples are given below.

%%% The lists of data figures and data tables are created automatically
%%% by the \listofDfigures and \listofDtables commands.

\newpage

\tableofcontents
\listofDtables
\listofDfigures
\vskip5pc

%%%% Authors begin text of article here %%%

\section{Introduction}

Ne-like ions occur in a variety of plasmas, such as astrophysical, laser-produced, magnetically confined, and Z-pinch -- see for example \cite{hag} and references therein.  Particularly important among these are the iron group of elements (namely Ti, Cr, Fe and Ni), whose lines are highly useful for the modelling and diagnostics of  a range of plasmas, including astrophysical, fusion and laser generated. However, to interpret observations or to model the plasmas,  information is required for several atomic parameters, such as energy levels, radiative rates and collision strengths. Therefore, over a period of time we have reported atomic data for a few Ne-like ions, namely Cl~VIII \cite{cl8}, Fe~XVII \cite{fe17a, fe17b}, Ni~XIX \cite{ni19a, ni19b}, and very recently W~LXV \cite{w65}. Similarly, many other workers have reported data for Ne-like ions. The most recent works are by J\"{o}nsson et al. \cite{jon} and Singh and Aggarwal \cite{sin}, who have also provided references for earlier works.

The majority of calculations available in the literature are confined to the lowest 27 levels of the 2s$^2$2p$^6$ and 2s$^2$2p$^5$3$\ell$ configurations, for which those by J\"{o}nsson et al. \cite{jon} are probably the most accurate, because differences between the measured and calculated energy levels are minimal. Moreover, they have covered a wider range of ions with 12 $\le$ Z $\le$ 36. However, very recently Singh and Aggarwal \cite{sin} (henceforth to be referred to as SA) have reported energy levels, radiative rates (A-values) and lifetimes ($\tau$) for three Ne-like ions, namely Cu~XX, Zn~XXI and Ga~XXII, among 127 levels of the 2s$^2$2p$^6$, 2s$^2$2p$^5$$3\ell$, 2s2p$^6$$3\ell$,  2s$^2$2p$^5$4$\ell$,  2s2p$^6$$4\ell$, and 2s$^2$2p$^5$5$\ell$ ($\ell \le$ 3) configurations.  We are not aware of any astrophysical importance of these ions, but they are useful for lasing and fusion plasmas \cite{elt, pals1, pals2}. For the calculations, they have adopted two independent atomic structure codes, namely  the general-purpose relativistic atomic structure package (GRASP) and  the flexible atomic code (FAC) of Gu \cite{fac}. Both of these codes are freely available on the websites {\tt http://amdpp.phys.strath.ac.uk/UK\_APAP/codes.html} and {\tt https://www-amdis.iaea.org/FAC/}, respectively. It was mainly done for the assessment of accuracy, particularly for the energy levels, because prior similar data, experimental or theoretical,  for most of the levels for these three ions do not exist. Therefore, based on the two calculations they concluded that both codes provide `comparable energies'. However, we notice that for many levels of all three ions the differences between the two sets of energies are significant, i.e. up to 1.5~Ryd -- see for example, levels 111--127 of Cu~XX, 112--119 of Zn~XXI and 111-116 of Ga~XXII in their tables 1--3.  In our long experience for a wide range of ions we have not noticed such large differences between calculations with these two codes, particularly when the same level of CI (configuration interaction) has been used. Additionally, in the absence of any other similar data, it is difficult to know which set of energies is more accurate. Apart from this, there are other reasons to perform yet another calculation, as discussed below.

The 127 levels considered by SA \cite{sin} are not the lowest in energy, because some from 2s$^2$2p$^5$5g intermix. They have included limited CI, among 51 configurations, namely 2s$^2$2p$^6$, 2s$^2$2p$^5$$n\ell$ (3 $\le n \le$ 7, but $\ell \le$ 3), 2s2p$^6$$n\ell$ (3 $\le n \le$ 7, but  $\ell \le$ 3), 2s$^2$2p$^4$3$\ell$3$\ell'$,  and 2s$^2$2p$^4$3$\ell$3$\ell'$. These configurations generate 1016 levels in total, but there is scope for improvement by including additional CI, particularly from the neglected configurations with $\ell >$ 3. More importantly, they have reported A-values only for the resonance transitions, i.e. from the ground level alone, whereas for accurate modelling of plasmas data are required for {\em all} transitions. Finally, for the reliability and accuracy of atomic data, which can be confidently applied, multiple calculations are preferred as emphasised by \cite{rev1, rev2}.

\section{Energy levels}

In our calculations we adopt the same GRASP0 version as by SA \cite{sin}. It was originally developed by Grant et al. \cite{grasp0} but has been considerably improved and updated by one of the authors, i.e. P.H.~Norrington. For the optimisation of the orbitals we  use the option of  `extended average level' (EAL),  in which a weighted (proportional to 2$j$+1) trace of the Hamiltonian matrix is minimised. The contributions of higher relativistic operators, namely Breit and quantum electrodynamic effects (QED), are also included. Although these effects are more  important for the heavy ions, their contributions improve the accuracy  of calculated results as may be noted from tables 1--3 of \cite{sin}. For the calculations, we have gradually increased the CI and our final ones have been performed with 64 configurations, namely 2s$^2$2p$^6$, 2s$^2$2p$^5$$3\ell$, 2s2p$^6$$3\ell$,  2s$^2$2p$^5$4$\ell$,  2s2p$^6$$4\ell$,  2s$^2$2p$^5$5$\ell$, 2s2p$^6$$5\ell$, 2s$^2$2p$^5$6s/p/d, 2s$^2$2p$^5$7s/p/d, (2s$^2$2p$^4$) 3s3p, 3s3d, 3p3d, 3s$^2$, 3p$^2$, 3d$^2$, 3s4$\ell$, 3s5$\ell$, 3p4$\ell$, 3p5$\ell$, 3p5$\ell$, 3d4$\ell$, and 3d5$\ell$, which generate 3948 levels (or configuration state functions, CSF) in total. Inclusion of further CI has not been possible in the version of the GRASP code adopted here, but is definitely possible though other versions, particularly GRASP2K \cite{grasp2k}, as has been done by J\"{o}nsson et al. \cite{jon}. However, we will assess the effect of additional CI by other means, as discussed below.

For the assessment of accuracy of energy levels and to gauge the significance (if any) of additional CI, we have performed analogous calculations with FAC. This  is also a fully relativistic code and the past experience (on a wide range of ions) shows that the calculated energy levels are generally in (close) agreement with those with GRASP or any other atomic structure code. Apart from this it has an added advantage of efficiency and the possibility of including a very (very) large CI. Therefore, we have performed  three calculations with increasing CI, which are: (i) FAC1, which includes 1112 levels of (almost) the  same configurations as included by SA \cite{sin}, i.e. 2*8, (2*7) 3*1, 4*1, 5*1, 6*1, 7*1, (2s2p$^5$) 3*2, and (2s$^2$2p$^4$) 3*2,  (ii) FAC2, which includes 17~729 levels arising from all possible combinations of  2*8, (2*7) 3*1, 4*1, 5*1, 6*1, 7*1, (2*6) 3*2, 3*1 4*1, 3*1 5*1, 3*1 6*1, and 3*1 7*1,  and finally (iii) FAC3, which includes a total of 93~437 levels, the additional ones arising from  (2*6) 4*1 5*1, 4*1 6*1, 4*1 7*1, 5*1 6*1, 5*1 7*1, 6*1 7*1, and 2*5 3*3. As an example, the energy spans of the levels in these calculations for Zn~XXI are 190, 240 and 290~Ryd, respectively, whereas in GRASP is up to 212~Ryd.

Since, as stated above, most of the calculations (and measurements) are confined to the lowest 27 levels of the 2s$^2$2p$^6$ and 2s$^2$2p$^5$3$\ell$ configurations, we make detailed comparisons for these in Tables~A, B and C for Cu~XX, Zn~XXI and Ga~XXII, respectively. Included in these tables are the experimental energies,  compiled by the NIST ((National Institute of Standards and Technology)) team from various sources, and  available at  the  website \\
{\tt http://www.nist.gov/pml/data/asd.cfm}, earlier calculations of J\"{o}nsson et al. \cite{jon} and SA \cite{sin}, and our work with GRASP and FAC. Measurements are available for most levels of Cu~XX \cite{nist}, but for only a few for other ions. It is clear from these tables that the calculated energies of SA with the GRASP and FAC codes do not match for any ion under consideration, and differ by up to $\sim$0.08~Ryd. These differences are not very significant and are often observed between any two calculations, but what is important is that their  energies with FAC differ more (by up to $\sim$0.2~Ryd) with our calculations, i.e. their results cannot be reproduced.  It may be worth noting here that  their calculations with both the GRASP and FAC codes had similar discrepancies  in the past, for a range of ions,  see for example the energy levels of five Br-like ions \cite{brlike} with 38 $\le$ Z $\le$ 42 and F-like W~LXVI \cite{w66a, w66b}.

Energies obtained by SA \cite{sin} with the GRASP code are comparable with our similar calculations with the {\em same} CI, but  not listed in Tables~A, B and C. However, the effect of additional CI included in our present calculations with GRASP is up to $\sim$0.1~Ryd, on the lowest 27 levels of Ne-like ions. Almost for all levels of the three ions, our energies are consistently higher than those of SA, and hence agree more closely with those of NIST and J\"{o}nsson et al. \cite{jon}. However, differences between the NIST and theoretical energies of J\"{o}nsson et al. are insignificant, but are up to $\sim$0.08~Ryd with our calculations with GRASP, and the latter are invariably lower. This may be partly due to the corresponding differences in the calculations of ground level energies and partly due to the larger CI included by J\"{o}nsson et al. On the other hand, our energies obtained with FAC are invariably {\em lower} than all other theoretical and experimental results, and an inclusion of large(r) CI has an insignificant effect (of less than 0.05~Ryd), and the discrepancies with those of NIST are up to 0.25~Ryd. Therefore, for the levels and the ions discussed here an inclusion of larger CI in the FAC calculations has been of no advantage, and what has been included in the GRASP  appears to be sufficient  for the accurate and reliable determination of energy levels. 

A notable {\em exception} in Tables~A, B and C is the 2s$^2$2p$^5$3d~$^1$P$^o_{1}$ level of Zn~XXI for which all theoretical energies are {\em lower} than of NIST,  by up to 0.45~Ryd. We have performed several other calculations with differing amount of CI, with both GRASP and FAC, but are unable to get an energy closer to the NIST value. Therefore,  J\"{o}nsson et al. \cite{jon} speculated a possibility of error in the determination of this measured result and suggested re-measurement. However, a closer examination of the original papers on which the NIST compilations are based reveals that for the concerned level, the measurement has been performed first by Burkhalter et al. \cite{bnc} and later by Boiko et al. \cite{bfp} from laser produced plasma sources, and they both reported $\lambda$ for the 3C (2s$^2$2p$^6$~$^1$S$_0$ -- 2s$^2$2p$^5$3d~$^1$P$^o_1$) transition to be 10.462 \AA\, with an accuracy of $\pm$2~m\AA, or equivalently an energy of 87.103 ($\pm$0.016)~Ryd, which {\em agrees excellently} with our result with GRASP or that of J\"{o}nsson et al. with GRASP2K.  Additionally, based on the Cowan's code \cite{cow} they calculated $\lambda$ to be 10.47 and 10.459~\AA, i.e. 87.036 and 87.128~Ryd, respectively.  This further confirms  that there is no discrepancy between theory and measurement for this level.  This anomaly has also been confirmed by one of the NIST team member (Alexander Kramida) and will (hopefully) be corrected soon on their website. 

For higher lying levels the only results available in the literature with which to compare are those by SA \cite{sin} for all three ions, and by Abdelaziz et al. \cite{abdel} for Cu~XX who have adopted the Cowan's code \cite{cow} based on Hartree-Fock-Slatter method. They have included 65 levels of the 2s$^2$2p$^6$, 2s$^2$2p$^5$$3\ell$ and  2s$^2$2p$^5$$4\ell$ configurations, but have neglected those from 2s2p$^6$$3\ell$, which intermix. Additionally, several of their levels are non-degenerate in energy -- see for example, 28/29, 41/42/43, 52/53, and 63/64 in their table~4. Therefore, we will not make comparisons with their results. Similarly, the results obtained by SA with FAC are not correct, as already stated, and those obtained with GRASP are comparable with our similar calculations with {\em limited} CI. Since the energies obtained in our present calculations with a larger CI are more accurate, as partly demonstrated in Tables~A, B and C, we list these in Tables~1--3 for all three ions. Also included in these tables, for comparison purposes, are the results with our largest calculations with FAC. The 2s$^2$2p$^6$, 2s$^2$2p$^5$$3\ell$, 2s2p$^6$$3\ell$,  2s$^2$2p$^5$4$\ell$,  2s2p$^6$$4\ell$, 2s$^2$2p$^5$5$\ell$, and  2s2p$^6$$5\ell$ (25) configurations generate 157 levels in total, but their energies are {\em not} the lowest, because some from other configurations intermix, such as of 2s$^2$2p$^5$6$\ell$. Therefore, we are restricting our results to the lowest 139 for which levels of other configurations do not mix.

As is often the case for most calculations, the LSJ$^{\pi}$ designations provided in Tables~1--3 are not unique for all levels, because for a few the eigenvector from a particular level/configuration dominates for more than one. This has been discussed several times in the past, by many workers. Therefore,  to help comparisons with other calculations sometimes mixing coefficients are listed, as has been done by J\"{o}nsson et al. \cite{jon} and SA \cite{sin}. Since we too have similar compositions, we are not listing these in Tables~1--3, but the results can be provided by the author on request. Additionally, the level orderings from GRASP and FAC are (generally) similar with only a few exceptions, but it is not always possible to find a perfect correspondence between the two calculations, because of different levels of CI included and the designations (nomenclatures) provided by the two codes. Nevertheless, in magnitude both calculations agree within about 0.2~Ryd with the FAC energies being lower, as was the case for the lowest 27 levels in Tables~A, B and C. In spite of using a larger CI in the FAC calculations our results with GRASP are expected to be (comparatively) more accurate, and this conclusion is based on the detailed comparisons discussed earlier for the lowest 27 levels, for all three ions.

 \section{Radiative rates}\label{sec.eqs} 

In Tables 4--6 we list our calculated results  with the GRASP code  for transition energies (wavelengths, $\lambda_{ji}$ in ${\rm \AA}$), radiative rates (A-values, in s$^{-1}$), oscillator strengths (f-values, dimensionless), and line strengths (S-values, in atomic units, 1 a.u. = 6.460$\times$10$^{-36}$ cm$^2$ esu$^2$)  for  electric dipole (E1) transitions in Ne-like ions Cu~XX, Zn~XXI and Ga~XXII.  These results are listed among the same 139 levels given in Tables~1--3. Although results have been  obtained in both the  velocity (Coulomb) and length (Babushkin) gauges, those from the latter are listed alone, because these are (generally) considered to be more accurate. However, their  ratio (R) is included in these tables. For a majority of  strong transitions (with f $\ge$ 0.01) R is close(r) to unity but deviations may be large for a few (very) weak ones. For (more) accurate modelling of plasmas and determination of lifetimes (to be discussed later) not only data for all E1 transitions are required, but also for other types, namely magnetic dipole (M1), electric quadrupole (E2) and magnetic quadrupole (M2). Therefore, included in these tables are our A-values for these transitions also and  the corresponding data for f- or S-values can be  obtained using Eqs. (1-5) given in \cite{fe17b}.   However,  for brevity only transitions from the lowest 5 to higher excited levels are listed in Tables~4--6, but  full tables in the ASCII format are available online in the electronic version.

In Table~D we compare our f-values with those of J\"{o}nsson et al. \cite{jon} for all E1 transitions, but from the lowest 5 levels. This will give an idea about the accuracy of the results obtained. For most transitions, particularly with f $\ge$ 0.01, the two sets of calculations agree within 10\%. This is highly satisfactory and encouraging, and confirms, once more, that small differences in energy levels do not adversely affect the calculations of A-values. However, differences for a few (very) weak transitions, such as 3--13 and 5--7, are up to 25\%. It is very difficult to assess with confidence the accuracy of such weak transitions (f $\le$ 10$^{-4}$) because some variations in mixing coefficients can often make large differences due to additive or cancellation effects of their small magnitudes. J\"{o}nsson et al. have also reported A-values for M1, E2 and M2 transitions, for which there are no major discrepancies with our results. However, for a few transitions some discrepancies are unavoidable, and these are reflected in the determination of $\tau$, discussed below.

\section{Lifetimes}

The lifetime $\tau$ of a level $j$ can be determined from the A-values  as it is equal to 1.0/$\Sigma_{i}$A$_{ji}$, where the summation runs  over all types of transitions, i.e. E1, E2, M1, and M2.  It is a measurable quantity but to the best of our knowledge no experiments have yet been performed for transitions/levels of the ions under consideration. However, J\"{o}nsson et al. \cite{jon} and SA \cite{sin} have listed $\tau$  and in Table~E we make comparisons with their results for the lowest 27 levels of Cu~XX, Zn~XXI and Ga~XXII. As expected there are no significant discrepancies among the three sets of results, and the only  level which stands out is 2p$^5$3s~$^3$P$^o_2$ of Cu~XX for which our value is lower by a factor of two. The only transition contributing to this level is 1--2 M2 for which our A-value is 1.27$\times$10$^{6}$~s$^{-1}$, whereas of J\"{o}nsson et al.  and SA are 6.30$\times$10$^{5}$~s$^{-1}$, and 6.81$\times$10$^{5}$~s$^{-1}$, respectively,  and hence the difference. Since this is a very weak transition with f = 1.12$\times$10$^{-7}$, it cannot be said with confidence which result is more accurate. For future comparisons, with experimental or theoretical data, our calculated values of $\tau$ are included  in Tables 1--3, and based on the comparisons discussed here and earlier for the A-values, our results are expected to be accurate  $\sim$20\%, for most levels/transitions.

\section{Conclusions}

In this paper, energies for the lowest 139 levels among the 2s$^2$2p$^6$, 2s$^2$2p$^5$$3\ell$, 2s2p$^6$$3\ell$,  2s$^2$2p$^5$4$\ell$,  2s2p$^6$$4\ell$, 2s$^2$2p$^5$5$\ell$, and  2s2p$^6$$5\ell$  configurations  are reported for three Ne-like ions, Cu~XX, Zn~XXI and Ga~XXII. The GRASP code has been adopted for the calculations and a reasonably large CI among 64 configurations (3948 CSF) has been included, which has been found to be sufficient, based on comparisons with other available theoretical and experimental energy levels. Nevertheless, calculations have also been performed with much larger CI (including 93~437 CSF) with FAC  to assess the effect and the accuracy. However, inclusion of such a large CI is not found to be beneficial for the levels considered in the paper. Based on multiple comparisons, particularly among the lowest 27 levels, our energies are assessed to be accurate to within 0.1~Ryd, for all ions. 

Further calculations for A-values (and other related parameters) have also been performed for four types of transitions, namely E1, E2, M1, and M2. These results are also listed, for all transitions among the 139 levels, and hence are  significantly more extensive than currently  available in the literature.  Comparisons with the existing literature have been possible, mostly among the lowest 27 levels, for which  no major discrepancies are found. Based on the comparisons made with other calculations, for both A-values and $\tau$, our results are assessed   to be  accurate to $\sim$20\%. However, this assessment of accuracy applies mostly to the  strong transitions with large f-values. For weak(er) transitions the reported A-values may be less certain.

%You are welcome to use BiBTeX with the \adndtbst\ bibliography
%style distributed with \adndtstyle\ package. This style comes very close
%to the journal style. Be sure to provide your
%\texttt{.bbl}  file (not the \texttt{.bib} file) with your submission.

%Please see \adndtguide\ for more instructions.

\ack
We thank the anonymous referee whose simple observation prompted us to resolve the discrepancy between theory and measurement for one of the levels of Zn~XXI and Alexander Kramida who very graciously and promptly confirmed the error in the NIST listing. 

\begin{appendix}

\def\thesection{} % To get the appendix heading correct

\section{Appendix A. Supplementary data}% The Appendix itself

Owing to space limitations, only parts of Tables 4--6  are presented here, but full tables are being made available as supplemental material in conjunction with the electronic publication of this work. Supplementary data associated with this article can be found, in the online version, at doi:nn.nnnn/j.adt.2018.nn.nnn.

\end{appendix}

%%  All sections inside the appendix environment will be appendixes
%%  Subsections function normally in appendixes.
%\newpage

%\end{document}

%\vspace*{1.1 cm}
\newpage
\renewcommand{\baselinestretch}{1.0}
\footnotesize
\begin{longtable}{@{\extracolsep\fill}rllrrrrrrrrr@{}}
\caption{Comparison of energies (in Ryd) for the lowest 27 levels of  Cu XX.}
Index  &     Configuration 	& Level 	     &  NIST	  & GRASP2K   & GRASP1	   &   GRASP2    &    FAC1a    &  FAC1b     &    FAC2   &    FAC3  \\  \\
\hline \\
\endfirsthead\\
\caption[]{(continued)}
Index  &     Configuration 	& Level 	     &  NIST	  & GRASP2K   & GRASP1	   &   GRASP2    &    FAC1a    &  FAC1b     &    FAC2   &    FAC3 \\  \\
\hline \\

\endhead 
    1  &    2s$^2$2p$^6$    	&     $^1$S$  _{0}$  &   0.0000    &   0.0000  &   0.0000   &	 0.0000   &   0.0000	&   0.0000   &   0.0000  &   0.0000  \\
    2  &    2s$^2$2p$^5$3s  	&     $^3$P$^o_{2}$  &  70.8717    &  70.8537  &  70.6919   &	70.7742   &  70.7705	&  70.6375   &  70.6174  &  70.6035  \\
    3  &    2s$^2$2p$^5$3s  	&     $^1$P$^o_{1}$  &  71.0392    &  71.0219  &  70.8653   &	70.9440   &  70.9535	&  70.8031   &  70.7827  &  70.7694  \\
    4  &    2s$^2$2p$^5$3s  	&     $^3$P$^o_{0}$  &  72.3906    &  72.3830  &  72.2166   &	72.3024   &  72.2977	&  72.1562   &  72.1332  &  72.1195  \\
    5  &    2s$^2$2p$^5$3s  	&     $^3$P$^o_{1}$  &  72.4917    &  72.4797  &  72.3172   &	72.4005   &  72.4053	&  72.2515   &  72.2283  &  72.2150  \\
    6  &    2s$^2$2p$^5$3p  	&     $^3$S$  _{1}$  &  73.5454    &  73.5319  &  73.3742   &	73.4594   &  73.4435	&  73.3253   &  73.3125  &  73.2979  \\
    7  &    2s$^2$2p$^5$3p  	&     $^3$D$  _{2}$  &  73.7969    &  73.7833  &  73.6392   &	73.7167   &  73.7184	&  73.5810   &  73.5642  &  73.5510  \\
    8  &    2s$^2$2p$^5$3p  	&     $^3$D$  _{3}$  &  74.0458    &  74.0317  &  73.8841   &	73.9649   &  73.9592	&  73.8301   &  73.8134  &  73.7998  \\
    9  &    2s$^2$2p$^5$3p  	&     $^1$P$  _{1}$  &  74.1171    &  74.1027  &  73.9583   &	74.0400   &  74.0387	&  73.9021   &  73.8843  &  73.8710  \\
   10  &    2s$^2$2p$^5$3p  	&     $^3$P$  _{2}$  &  74.3009    &  74.2868  &  74.1445   &	74.2200   &  74.2258	&  74.0841   &  74.0666  &  74.0535  \\
   11  &    2s$^2$2p$^5$3p  	&     $^3$P$  _{0}$  &             &  74.8788  &  74.7588   &	74.8281   &  74.8408	&  74.6842   &  74.6676  &  74.6526  \\
   12  &    2s$^2$2p$^5$3p  	&     $^3$D$  _{1}$  &  75.2641    &  75.2555  &  75.1061   &	75.1907   &  75.1872	&  75.0452   &  75.0252  &  75.0119  \\
   13  &    2s$^2$2p$^5$3p  	&     $^3$P$  _{1}$  &  75.6456    &  75.6365  &  75.4856   &	75.5685   &  75.5648	&  75.4238   &  75.4043  &  75.3909  \\
   14  &    2s$^2$2p$^5$3p  	&     $^1$D$  _{2}$  &  75.6925    &  75.6833  &  75.5343   &	75.6157   &  75.6162	&  75.4697   &  75.4499  &  75.4368  \\
   15  &    2s$^2$2p$^5$3p  	&     $^1$S$  _{0}$  &  76.6090    &  76.6074  &  76.6747   &	76.6808   &  76.7670	&  76.4945   &  76.4922  &  76.4620  \\
   16  &    2s$^2$2p$^5$3d  	&     $^3$P$^o_{0}$  &  77.5448    &  77.5365  &  77.3901   &	77.4729   &  77.4471	&  77.3028   &  77.2885  &  77.2737  \\
   17  &    2s$^2$2p$^5$3d  	&     $^3$P$^o_{1}$  &  77.6474    &  77.6328  &  77.4886   &	77.5713   &  77.5468	&  77.4011   &  77.3862  &  77.3714  \\
   18  &    2s$^2$2p$^5$3d  	&     $^3$F$^o_{4}$  &  77.8230    &  77.8061  &  77.6721   &	77.7473   &  77.7263	&  77.5843   &  77.5682  &  77.5534  \\
   19  &    2s$^2$2p$^5$3d  	&     $^3$P$^o_{2}$  &  77.8290    &  77.8140  &  77.6720   &	77.7551   &  77.7326	&  77.5846   &  77.5692  &  77.5538  \\
   20  &    2s$^2$2p$^5$3d  	&     $^3$F$^o_{3}$  &  77.8681    &  77.8538  &  77.7222   &	77.8001   &  77.7778	&  77.6289   &  77.6116  &  77.5964  \\
   21  &    2s$^2$2p$^5$3d  	&     $^1$D$^o_{2}$  &  78.0302    &  78.0134  &  77.8832   &	77.9637   &  77.9411	&  77.7897   &  77.7704  &  77.7554  \\
   22  &    2s$^2$2p$^5$3d  	&     $^3$D$^o_{3}$  &  78.1367    &  78.1229  &  77.9900   &	78.0719   &  78.0527	&  77.8962   &  77.8772  &  77.8625  \\
   23  &    2s$^2$2p$^5$3d  	&     $^3$D$^o_{1}$  &  78.6105    &  78.5948  &  78.4870   &	78.5594   &  78.5407	&  78.3764   &  78.3560  &  78.3387  \\
   24  &    2s$^2$2p$^5$3d  	&     $^3$F$^o_{2}$  &  79.3815    &  79.3787  &  79.2339   &	79.3150   &  79.2971	&  79.1420   &  79.1217  &  79.1066  \\
   25  &    2s$^2$2p$^5$3d  	&     $^3$D$^o_{2}$  &  79.4660    &  79.4545  &  79.3136   &	79.3994   &  79.3733	&  79.2132   &  79.1927  &  79.1781  \\
   26  &    2s$^2$2p$^5$3d  	&     $^1$F$^o_{3}$  &  79.5283    &  79.5139  &  79.3771   &	79.4605   &  79.4361	&  79.2746   &  79.2538  &  79.2390  \\
   27  &    2s$^2$2p$^5$3d  	&     $^1$P$^o_{1}$  &  80.0731    &  80.0679  &  79.9972   &	80.0561   &  80.0433	&  79.8589   &  79.8345  &  79.8137  \\
\\  \hline            								                	 
\end{longtable}   								   					       
			      							   					       
%\vspace*{0.5 cm}													       

%}													       
														       
\begin{flushleft}													       
{\small
NIST:   Experimental energies compiled by the NIST team and available at the website {\tt http://www.nist.gov/pml/data/asd.cfm}  \\	
GRASP2K: Calculations of Jonsson et al. \cite{jon} with the GRASP2K code \\					      		      
GRASP1: Calculations of Singh and Aggarwal \cite{sin} with the GRASP code for 1016 levels \\
GRASP2: Present calculations  with the GRASP code for 3948 levels \\
FAC1a: Calculations of Singh and Aggarwal \cite{sin} with the FAC code for 1112 levels \\
FAC1b: Present calculations  with the FAC code  for 1112 levels \\
FAC2: Present calculations  with the FAC code  for 17~729 levels \\
FAC3: Present calculations  with the FAC code  for 93~437 levels \\   
}															       
\end{flushleft} 

%\end{document}

\clearpage
\newpage

\renewcommand{\baselinestretch}{1.0}
\footnotesize
\begin{longtable}{@{\extracolsep\fill}rlllrrrrrrrr@{}}
\caption{Comparison of energies (in Ryd) for the lowest 27 levels of  Zn XXI.}
Index  &     Configuration 	& Level 	     &  NIST	  & GRASP2K   & GRASP1	   &   GRASP2    &    FAC1a    &  FAC1b     &    FAC2   &    FAC3  \\  \\
\hline \\
\endfirsthead\\
\caption[]{(continued)}
Index  &     Configuration 	& Level 	     &  NIST	  & GRASP2K   & GRASP1	   &   GRASP2    &    FAC1a    &  FAC1b     &    FAC2   &    FAC3 \\  \\
\hline \\

\endhead 
    1  &    2s$^2$2p$^6$    	&     $^1$S$  _{0}$  &  00.000	  &   0.0000  &    0.0000  &    0.0000   &   0.0000    &   0.0000   &   0.0000  &  0.0000   \\
    2  &    2s$^2$2p$^5$3s  	&     $^3$P$^o_{2}$  &  	  &  77.2445  &   77.0829  &   77.1655   &  77.1622    &  77.0297   &  77.0111  & 76.9965   \\
    3  &    2s$^2$2p$^5$3s  	&     $^1$P$^o_{1}$  & 77.429	  &  77.4220  &   77.2657  &   77.3445   &  77.3546    &  77.2046   &  77.1858  & 77.1717   \\
    4  &    2s$^2$2p$^5$3s  	&     $^3$P$^o_{0}$  &  	  &  79.0279  &   78.8614  &   78.9477   &  78.9435    &  78.8018   &  78.7802  & 78.7657   \\
    5  &    2s$^2$2p$^5$3s  	&     $^3$P$^o_{1}$  & 79.131	  &  79.1277  &   78.9649  &   79.0489   &  79.0539    &  78.9002   &  78.8784  & 78.8643   \\
    6  &    2s$^2$2p$^5$3p  	&     $^3$S$  _{1}$  &  	  &  80.0759  &   79.9196  &   80.0046   &  79.9897    &  79.8713   &  79.8599  & 79.8446   \\
    7  &    2s$^2$2p$^5$3p  	&     $^3$D$  _{2}$  &  	  &  80.3236  &   80.1804  &   80.2582   &  80.2601    &  80.1232   &  80.1080  & 80.0941   \\
    8  &    2s$^2$2p$^5$3p  	&     $^3$D$  _{3}$  &  	  &  80.6297  &   80.4828  &   80.5639   &  80.5583    &  80.4299   &  80.4147  & 80.4004   \\
    9  &    2s$^2$2p$^5$3p  	&     $^1$P$  _{1}$  &  	  &  80.6955  &   80.5515  &   80.6338   &  80.6322    &  80.4966   &  80.4804  & 80.4663   \\
   10  &    2s$^2$2p$^5$3p  	&     $^3$P$  _{2}$  & 80.930	  &  80.8980  &   80.7564  &   80.8322   &  80.8381    &  80.6971   &  80.6811  & 80.6674   \\
   11  &    2s$^2$2p$^5$3p  	&     $^3$P$  _{0}$  &  	  &  81.5606  &   81.4461  &   81.5142   &  81.5289    &  81.3700   &  81.3555  & 81.3392   \\
   12  &    2s$^2$2p$^5$3p  	&     $^3$D$  _{1}$  &  	  &  82.0446  &   81.8956  &   81.9807   &  81.9773    &  81.8353   &  81.8167  & 81.8027   \\
   13  &    2s$^2$2p$^5$3p  	&     $^3$P$  _{1}$  &  	  &  82.4876  &   82.3366  &   82.4203   &  82.4164    &  82.2758   &  82.2578  & 82.2436   \\
   14  &    2s$^2$2p$^5$3p  	&     $^1$D$  _{2}$  & 82.568	  &  82.5407  &   82.3920  &   82.4740   &  82.4746    &  82.3283   &  82.3098  & 82.2960   \\
   15  &    2s$^2$2p$^5$3p  	&     $^1$S$  _{0}$  & 83.453	  &  83.4324  &   83.4967  &   83.5049   &  83.5888    &  83.3206   &  83.3191  & 83.2879   \\
   16  &    2s$^2$2p$^5$3d  	&     $^3$P$^o_{0}$  &  	  &  84.3033  &   84.1579  &   84.2407   &  84.2146    &  84.0717   &  84.0589  & 84.0433   \\
   17  &    2s$^2$2p$^5$3d  	&     $^3$P$^o_{1}$  & 84.408	  &  84.4092  &   84.2661  &   84.3490   &  84.3240    &  84.1796   &  84.1663  & 84.1506   \\
   18  &    2s$^2$2p$^5$3d  	&     $^3$F$^o_{4}$  &  	  &  84.5963  &   84.4631  &   84.5385   &  84.5172    &  84.3768   &  84.3630  & 84.3468   \\
   19  &    2s$^2$2p$^5$3d  	&     $^3$P$^o_{2}$  &  	  &  84.6078  &   84.4671  &   84.5502   &  84.5274    &  84.3802   &  84.3656  & 84.3500   \\
   20  &    2s$^2$2p$^5$3d  	&     $^3$F$^o_{3}$  &  	  &  84.6385  &   84.5079  &   84.5860   &  84.5635    &  84.4157   &  84.4000  & 84.3840   \\
   21  &    2s$^2$2p$^5$3d  	&     $^1$D$^o_{2}$  &  	  &  84.8102  &   84.6811  &   84.7618   &  84.7391    &  84.5887   &  84.5709  & 84.5552   \\
   22  &    2s$^2$2p$^5$3d  	&     $^3$D$^o_{3}$  &  	  &  84.9321  &   84.8000  &   84.8822   &  84.8624    &  84.7074   &  84.6899  & 84.6745   \\
   23  &    2s$^2$2p$^5$3d  	&     $^3$D$^o_{1}$  & 85.460	  &  85.4497  &   85.3457  &   85.4174   &  85.3986    &  85.2347   &  85.2157  & 85.1973   \\
   24  &    2s$^2$2p$^5$3d  	&     $^3$F$^o_{2}$  &  	  &  86.4039  &   86.2698  &   86.3512   &  86.3332    &  86.1788   &  86.1599  & 86.1440   \\
   25  &    2s$^2$2p$^5$3d  	&     $^3$D$^o_{2}$  &  	  &  86.5016  &   86.3610  &   86.4473   &  86.4209    &  86.2615   &  86.2424  & 86.2270   \\
   26  &    2s$^2$2p$^5$3d  	&     $^1$F$^o_{3}$  &  	  &  86.5680  &   86.4317  &   86.5155   &  86.4909    &  86.3300   &  86.3106  & 86.2950   \\
   27  &    2s$^2$2p$^5$3d  	&     $^1$P$^o_{1}$  & 87.340$^{\rm a}$	  &  87.1303  &   87.0583  &   87.1188   &  87.1054    &  86.9226   &  86.8997  & 86.8781   \\

\\  \hline            								                	 
\end{longtable}   								   					       
			      							   					       
%\vspace*{0.5 cm}													       

%}													       
														       
\begin{flushleft}													       
{\small
NIST:   Experimental energies compiled by the NIST team and available at the website  {\tt http://www.nist.gov/pml/data/asd.cfm}  \\	
GRASP2K: Calculations of Jonsson et al. \cite{jon} with the GRASP2K code \\					      		      
GRASP1: Calculations of Singh and Aggarwal \cite{sin}  with the GRASP code for 1016 levels \\
GRASP2: Present calculations  with the GRASP code for 3948 levels \\
FAC1a: Calculations of Singh and Aggarwal \cite{sin} with the FAC code for 1112 levels \\
FAC1b: Present calculations  with the FAC code  for 1112 levels \\
FAC2: Present calculations  with the FAC code  for 17~729 levels \\
FAC3: Present calculations  with the FAC code  for 93~437 levels \\   
a: see text in Section 2\\
}															       
\end{flushleft} 

%\end{document}

\clearpage
\newpage

\renewcommand{\baselinestretch}{1.0}
\footnotesize
\begin{longtable}{@{\extracolsep\fill}rllrrrrrrrrr@{}}
\caption{Comparison of energies (in Ryd) for the lowest 27 levels of  Ga XXII.}
Index  &     Configuration 	& Level 	     &  NIST	  & GRASP2K   & GRASP1	   &   GRASP2    &    FAC1a    &  FAC1b     &    FAC2   &    FAC3  \\  \\
\hline \\
\endfirsthead\\
\caption[]{(continued)}
Index  &     Configuration 	& Level 	     &  NIST	  & GRASP2K   & GRASP1	   &   GRASP2    &    FAC1a    &  FAC1b     &    FAC2   &    FAC3 \\  \\
\hline \\

\endhead 
    1  &    2s$^2$2p$^6$    	&     $^1$S$  _{0}$  &   0.000    &   0.0000  &   0.0000   &   0.0000	 &   0.0000    &   0.0000   &	0.0000  &   0.0000  \\
    2  &    2s$^2$2p$^5$3s  	&     $^3$P$^o_{2}$  &            &  83.9042  &  83.7426   &  83.8255	 &  83.8227    &  83.6908   &  83.6736  &  83.6583  \\
    3  &    2s$^2$2p$^5$3s  	&     $^1$P$^o_{1}$  &  84.090    &  84.0909  &  83.9348   &  84.0138	 &  84.0246    &  83.8748   &  83.8575  &  83.8428  \\
    4  &    2s$^2$2p$^5$3s  	&     $^3$P$^o_{0}$  &            &  85.9727  &  85.8058   &  85.8926	 &  85.8890    &  85.7471   &  85.7268  &  85.7116  \\
    5  &    2s$^2$2p$^5$3s  	&     $^3$P$^o_{1}$  &  86.100    &  86.0756  &  85.9122   &  85.9968	 &  86.0023    &  85.8487   &  85.8282  &  85.8134  \\
    6  &    2s$^2$2p$^5$3p  	&     $^3$S$  _{1}$  &            &  86.8900  &  86.7349   &  86.8197	 &  86.8059    &  86.6873   &  86.6773  &  86.6613  \\
    7  &    2s$^2$2p$^5$3p  	&     $^3$D$  _{2}$  &            &  87.1336  &  86.9911   &  87.0691	 &  87.0714    &  86.9350   &  86.9213  &  86.9067  \\
    8  &    2s$^2$2p$^5$3p  	&     $^3$D$  _{3}$  &            &  87.5057  &  87.3592   &  87.4406	 &  87.4354    &  87.3076   &  87.2939  &  87.2788  \\
    9  &    2s$^2$2p$^5$3p  	&     $^1$P$  _{1}$  &            &  87.5659  &  87.4221   &  87.5050	 &  87.5033    &  87.3686   &  87.3540  &  87.3392  \\
   10  &    2s$^2$2p$^5$3p  	&     $^3$P$  _{2}$  &  87.786    &  87.7871  &  87.6461   &  87.7222	 &  87.7284    &  87.5880   &  87.5735  &  87.5591  \\
   11  &    2s$^2$2p$^5$3p  	&     $^3$P$  _{0}$  &            &  88.5231  &  88.4146   &  88.4812	 &  88.4982    &  88.3369   &  88.3243  &  88.3068  \\
   12  &    2s$^2$2p$^5$3p  	&     $^3$D$  _{1}$  &            &  89.1342  &  88.9853   &  89.0710	 &  89.0679    &  88.9259   &  88.9087  &  88.8939  \\
   13  &    2s$^2$2p$^5$3p  	&     $^3$P$  _{1}$  &            &  89.6493  &  89.4964   &  89.5807	 &  89.5769    &  89.4364   &  89.4199  &  89.4049  \\
   14  &    2s$^2$2p$^5$3p  	&     $^1$D$  _{2}$  &  89.725    &  89.7069  &  89.5583   &  89.6410	 &  89.6419    &  89.4955   &  89.4783  &  89.4638  \\
   15  &    2s$^2$2p$^5$3p  	&     $^1$S$  _{0}$  &            &  90.5554  &  90.6161   &  90.6266	 &  90.7082    &  90.4443   &  90.4436  &  90.4115  \\
   16  &    2s$^2$2p$^5$3d  	&     $^3$P$^o_{0}$  &            &  91.3482  &  91.2036   &  91.2865	 &  91.2602    &  91.1186   &  91.1073  &  91.0909  \\
   17  &    2s$^2$2p$^5$3d  	&     $^3$P$^o_{1}$  &  91.470    &  91.4642  &  91.3219   &  91.4048	 &  91.3798    &  91.2365   &  91.2246  &  91.2082  \\
   18  &    2s$^2$2p$^5$3d  	&     $^3$F$^o_{4}$  &            &  91.6664  &  91.5338   &  91.6093	 &  91.5879    &  91.4488   &  91.4365  &  91.4195  \\
   19  &    2s$^2$2p$^5$3d  	&     $^3$P$^o_{2}$  &            &  91.6808  &  91.5411   &  91.6243	 &  91.6013    &  91.4552   &  91.4420  &  91.4257  \\
   20  &    2s$^2$2p$^5$3d  	&     $^3$F$^o_{3}$  &            &  91.7010  &  91.5713   &  91.6495	 &  91.6270    &  91.4803   &  91.4661  &  91.4493  \\
   21  &    2s$^2$2p$^5$3d  	&     $^1$D$^o_{2}$  &            &  91.8856  &  91.7574   &  91.8384	 &  91.8157    &  91.6663   &  91.6498  &  91.6334  \\
   22  &    2s$^2$2p$^5$3d  	&     $^3$D$^o_{3}$  &            &  92.0212  &  91.8898   &  91.9722	 &  91.9525    &  91.7983   &  91.7823  &  91.7662  \\
   23  &    2s$^2$2p$^5$3d  	&     $^3$D$^o_{1}$  &  92.600    &  92.5850  &  92.4846   &  92.5555	 &  92.5370    &  92.3733   &  92.3557  &  92.3361  \\
   24  &    2s$^2$2p$^5$3d  	&     $^3$F$^o_{2}$  &            &  93.7480  &  93.6143   &  93.6960	 &  93.6780    &  93.5243   &  93.5067  &  93.4901  \\
   25  &    2s$^2$2p$^5$3d  	&     $^3$D$^o_{2}$  &            &  93.8595  &  93.7190   &  93.8057	 &  93.7793    &  93.6204   &  93.6027  &  93.5866  \\
   26  &    2s$^2$2p$^5$3d  	&     $^1$F$^o_{3}$  &            &  93.9328  &  93.7968   &  93.8810	 &  93.8564    &  93.6960   &  93.6780  &  93.6617  \\
   27  &    2s$^2$2p$^5$3d  	&     $^1$P$^o_{1}$  &  94.500    &  94.5004  &  94.4270   &  94.4890	 &  94.4753    &  94.2939   &  94.2724  &  94.2501  \\
\\  \hline            								                	 
\end{longtable}   								   					       
			      							   					       
%\vspace*{0.5 cm}													       

%}													       
														       
\begin{flushleft}													       
{\small
NIST:  Experimental energies compiled by the NIST team and available at the website   {\tt http://www.nist.gov/pml/data/asd.cfm}  \\	
GRASP2K: Calculations of Jonsson et al. \cite{jon} with the GRASP2K code \\					      		      
GRASP1: Calculations of Singh and Aggarwal \cite{sin} with the GRASP code for 1016 levels \\
GRASP2: Present calculations  with the GRASP code for 3948 levels \\
FAC1a: Calculations of Singh and Aggarwal \cite{sin} with the FAC code for 1112 levels \\
FAC1b: Present calculations  with the FAC code  for 1112 levels \\
FAC2: Present calculations  with the FAC code  for 17~729 levels \\
FAC3: Present calculations  with the FAC code  for 93~437 levels \\   
}															       
\end{flushleft} 

%\end{document}

\clearpage
\newpage

\renewcommand{\baselinestretch}{1.0}
\footnotesize
\begin{longtable}{@{\extracolsep\fill}rrrrrrrr@{}}
\caption{Comparison of oscillator strengths (f-values) for a few E1 transitions of Cu XX, Zn~XXI and Ga XXII. $a{\pm}b \equiv$ $a\times$10$^{{\pm}b}$.}

Ion & &  \multicolumn{2}{c}{Cu XX} & \multicolumn{2}{c}{Zn XXI} &\multicolumn{2}{c}{Ga XXII}   \\  \\
\hline  \\
I & J & GRASP & GRASP2K   & GRASP & GRASP2K   & GRASP & GRASP2K   \\  \\
 \hline  \\
\endfirsthead\\
\caption[]{(continued)}
Ion & &  \multicolumn{2}{c}{Cu XX} & \multicolumn{2}{c}{Zn XXI} &\multicolumn{2}{c}{Ga XXII}   \\  \\
\hline  \\
I & J & GRASP & GRASP2K   & GRASP & GRASP2K   & GRASP & GRASP2K   \\  \\
\hline  \\
\endhead 
1  &    3   &  1.31-1  & 1.28-1   &   1.32-1   &   1.29-1   &  1.33-1  & 1.30-1   \\  
1  &    5   &  9.53-2  & 9.20-2   &   9.29-2   &   9.98-2   &  9.09-2  & 8.80-2   \\  
1  &   17   &  9.90-3  & 1.00-2   &   9.68-3   &   9.76-3   &  9.34-3  & 9.37-3   \\  
1  &   23   &  9.03-1  & 9.24-1   &   1.00-0   &   1.02-0   &  1.10-0  & 1.11-0   \\  
1  &   27   &  2.29-0  & 2.15-0   &   2.23-0   &   2.10-0   &  2.17-0  & 2.05-0   \\  
2  &    6   &  4.80-2  & 4.78-2   &   4.71-2   &   4.69-2   &  4.62-2  & 4.60-2   \\  
2  &    7   &  4.58-2  & 4.52-2   &   4.44-2   &   4.38-2   &  4.30-2  & 4.25-2   \\  
2  &    8   &  1.48-1  & 1.46-1   &   1.44-1   &   1.43-1   &  1.41-1  & 1.40-1   \\  
2  &    9   &  1.08-3  & 9.63-4   &   5.63-4   &   4.83-4   &  2.22-4  & 1.74-4   \\  
2  &   10   &  6.23-2  & 6.13-2   &   6.05-2   &   5.98-2   &  5.90-2  & 5.81-2   \\  
2  &   12   &  1.17-4  & 1.16-4   &   1.19-4   &   1.17-4   &  1.18-4  & 1.15-4   \\  
2  &   13   &  3.24-3  & 3.05-3   &   2.86-3   &   2.68-3   &  2.53-3  & 2.38-3   \\  
2  &   14   &  5.61-4  & 5.35-4   &   4.65-4   &   4.44-4   &  3.90-4  & 3.72-4   \\  
3  &    6   &  7.42-4  & 7.01-4   &   2.86-4   &   2.60-4   &  4.93-5  & 3.88-5   \\  
3  &    7   &  8.10-2  & 8.00-2   &   7.75-2   &   7.66-2   &  7.44-2  & 7.35-2   \\  
3  &    9   &  9.97-2  & 9.85-2   &   9.76-2   &   9.65-2   &  9.56-2  & 9.46-2   \\  
3  &   10   &  8.64-2  & 8.52-2   &   8.49-2   &   8.38-2   &  8.36-2  & 8.25-2   \\  
3  &   11   &  3.38-2  & 3.43-2   &   3.43-2   &   3.47-2   &  3.48-2  & 3.52-2   \\  
3  &   12   &  2.85-4  & 2.40-4   &   2.46-4   &   2.09-4   &  2.15-4  & 1.83-4   \\  
3  &   13   &  8.52-5  & 6.55-5   &   9.11-5   &   7.22-5   &  9.45-5  & 7.70-5   \\  
3  &   14   &  6.48-4  & 6.14-4   &   5.08-4   &   4.83-4   &  4.04-4  & 3.85-4   \\  
3  &   15   &  2.12-2  & 1.87-2   &   1.94-2   &   1.71-2   &  1.77-2  & 1.55-2   \\  
4  &    6   &  7.97-4  & 8.04-4   &   5.42-4   &   5.47-4   &  3.54-4  & 3.57-4   \\  
4  &    9   &  9.88-5  & 9.16-5   &   4.78-5   &   4.36-5   &  2.14-5  & 1.91-5   \\  
4  &   12   &  1.04-1  & 1.01-1   &   9.80-2   &   9.58-2   &  9.29-2  & 9.10-2   \\  
4  &   13   &  2.01-1  & 2.00-1   &   1.99-1   &   1.98-1   &  1.96-1  & 1.95-1   \\  
5  &    6   &  3.50-4  & 3.51-4   &   2.36-4   &   2.37-4   &  1.51-4  & 1.52-4   \\  
5  &    7   &  5.07-6  & 3.67-6   &   3.52-6   &   2.56-6   &  2.45-6  & 1.79-6   \\  
5  &    9   &  4.02-5  & 3.32-5   &   4.17-5   &   3.58-5   &  4.05-5  & 3.58-5   \\  
5  &   10   &  5.28-4  & 5.19-4   &   3.75-4   &   3.70-4   &  2.66-4  & 2.61-4   \\  
5  &   11   &  6.82-3  & 6.04-3   &   5.83-3   &   5.11-3   &  4.89-3  & 4.26-3   \\  
5  &   12   &  5.71-2  & 5.67-2   &   5.55-2   &   5.51-2   &  5.40-2  & 5.35-2   \\  
5  &   13   &  3.57-2  & 3.50-2   &   3.43-2   &   3.37-2   &  3.31-2  & 3.25-2   \\  
5  &   14   &  1.78-1  & 1.76-1   &   1.74-1   &   1.72-1   &  1.71-1  & 1.69-1   \\  
5  &   15   &  3.96-2  & 3.87-2   &   3.86-2   &   3.78-2   &  3.78-2  & 3.70-2   \\
\\  \hline            								                	 
\end{longtable}   								   					       
			      							   					       
%\vspace*{0.5 cm}													       

%}													       
														       
\begin{flushleft}													       
{\small
GRASP: Present calculations  with the GRASP code   \\
GRASP2K: Calculations of J\"{o}nsson et al. \cite{jon} with the GRASP2K code \\      
}															       
\end{flushleft} 

%\end{document}

\clearpage
\newpage

\renewcommand{\baselinestretch}{1.0}
\footnotesize
\begin{longtable}{@{\extracolsep\fill}rlrrrrrrrrrrr@{}}
\caption{Comparison of lifetimes ($\tau$, s) for the lowest 27 levels of  Cu XX, Zn XXI and Ga XXII. $a{\pm}b \equiv$ $a\times$10$^{{\pm}b}$.}
Ion & & & \multicolumn{3}{c}{Cu XX} & \multicolumn{3}{c}{Zn XXI} &\multicolumn{3}{c}{Ga XXII}   \\  \\
\hline \\
Index  &     Config.  & Level  & GRASP$^a$ & GRASP$^b$ & GRASP$^c$ &  GRASP$^a$ & GRASP$^b$ & GRASP$^c$ &  GRASP$^a$ & GRASP$^b$ & GRASP$^c$  \\
\\ \hline  \\  
\endfirsthead\\
\caption[]{(continued)}
Ion & \multicolumn{3}{c}{Cu XX} & \multicolumn{3}{c}{Zn XXI} &\multicolumn{3}{c}{Ga XXII}   \\  \\
\hline \\
Index  &     Config.  & Level  & GRASP$^a$ & GRASP$^b$ & GRASP$^c$ &  GRASP$^a$ & GRASP$^b$ & GRASP$^c$ &  GRASP$^a$ & GRASP$^b$ & GRASP$^c$  \\
\\ \hline  \\  \\
\hline\\
\endhead 
    1   & 2p$^6$    &	$^1$S$  _{0}$	  &   ........  & ........   & .......    &  ........  & ........   &	.......  &   ........  & ........   &  .......     \\
    2   & 2p$^5$3s  &	$^3$P$^o_{2}$	  &   7.896-07  & 1.588-06   & 1.47-06    &  1.089-06  & 1.133-06   &	1.05-06  &   7.896-07  & 8.203-07   &  7.64-07     \\ 
    3   & 2p$^5$3s  &	$^1$P$^o_{1}$	  &   5.661-13  & 5.796-13   & 5.54-13    &  4.724-13  & 4.832-13   &	4.63-13  &   3.980-13  & 4.065-13   &  3.90-13     \\
    4   & 2p$^5$3s  &	$^3$P$^o_{0}$	  &   1.342-05  & 1.338-05   & 1.36-05    &  8.355-06  & 8.330-06   &	8.43-06  &   5.296-06  & 5.282-06   &  5.34-06     \\
    5   & 2p$^5$3s  &	$^3$P$^o_{1}$	  &   7.477-13  & 7.728-13   & 7.25-13    &  6.435-13  & 6.642-13   &	6.25-13  &   5.557-13  & 5.729-13   &  5.41-13     \\
    6   & 2p$^5$3p  &	$^3$S$  _{1}$	  &   2.138-10  & 2.157-10   & 2.15-10    &  1.958-10  & 1.977-10   &	1.97-10  &   1.802-10  & 1.821-10   &  1.81-10     \\
    7   & 2p$^5$3p  &	$^3$D$  _{2}$	  &   1.197-10  & 1.367-10   & 1.34-10    &  1.211-10  & 1.235-10   &	1.21-10  &   1.093-10  & 1.113-10   &  1.09-10     \\
    8   & 2p$^5$3p  &	$^3$D$  _{3}$	  &   1.159-10  & 1.182-10   & 1.16-10    &  1.046-10  & 1.066-10   &	1.05-10  &   9.440-11  & 9.619-11   &  9.45-11     \\
    9   & 2p$^5$3p  &	$^1$P$  _{1}$	  &   1.277-10  & 1.308-10   & 1.29-10    &  1.166-10  & 1.193-10   &	1.17-10  &   1.063-10  & 1.087-10   &  1.07-10     \\
   10   & 2p$^5$3p  &	$^3$P$  _{2}$	  &   7.921-11  & 8.700-11   & 8.51-11    &  7.555-11  & 7.708-11   &	7.55-11  &   6.693-11  & 6.823-11   &  6.69-11     \\
   11   & 2p$^5$3p  &	$^3$P$  _{0}$	  &   7.538-11  & 7.614-11   & 7.58-11    &  6.570-11  & 6.637-11   &	6.61-11  &   5.720-11  & 5.782-11   &  5.75-11     \\
   12   & 2p$^5$3p  &	$^3$D$  _{1}$	  &   1.678-10  & 1.720-10   & 1.69-10    &  1.581-10  & 1.620-10   &	1.59-10  &   1.493-10  & 1.529-10   &  1.50-10     \\
   13   & 2p$^5$3p  &	$^3$P$  _{1}$	  &   1.036-10  & 1.061-10   & 1.04-10    &  9.409-11  & 9.625-11   &	9.42-11  &   8.541-11  & 8.731-11   &  8.55-11     \\
   14   & 2p$^5$3p  &	$^1$D$  _{2}$	  &   8.613-11  & 9.614-11   & 9.43-11    &  8.319-11  & 8.474-11   &	8.31-11  &   7.329-11  & 7.460-11   &  7.32-11     \\
   15   & 2p$^5$3p  &	$^1$S$  _{0}$	  &   2.921-11  & 3.342-11   & 2.76-11    &  2.761-11  & 3.150-11   &	2.61-11  &   2.623-11  & 2.985-11   &  2.48-11     \\
   16   & 2p$^5$3d  &	$^3$P$^o_{0}$	  &   6.902-11  & 6.981-11   & 6.92-11    &  6.417-11  & 6.487-11   &	6.43-11  &   5.964-11  & 6.026-11   &  5.97-11     \\
   17   & 2p$^5$3d  &	$^3$P$^o_{1}$	  &   5.750-12  & 5.693-12   & 5.86-12    &  4.998-12  & 4.963-12   &	5.09-12  &   4.432-12  & 4.414-12   &  4.50-12     \\
   18   & 2p$^5$3d  &	$^3$F$^o_{4}$	  &   7.571-11  & 7.696-11   & 7.17-11    &  7.173-11  & 7.286-11   &	6.69-11  &   6.805-11  & 6.906-11   &  6.23-11     \\
   19   & 2p$^5$3d  &	$^3$P$^o_{2}$	  &   7.123-11  & 7.257-11   & 7.57-11    &  6.652-11  & 6.765-11   &	7.17-11  &   6.198-11  & 6.299-11   &  6.80-11     \\
   20   & 2p$^5$3d  &	$^3$F$^o_{3}$	  &   6.198-11  & 6.316-11   & 6.23-11    &  5.750-11  & 5.854-11   &	5.77-11  &   5.335-11  & 5.427-11   &  5.36-11     \\
   21   & 2p$^5$3d  &	$^1$D$^o_{2}$	  &   6.003-11  & 8.129-11   & 6.04-11    &  5.613-11  & 5.723-11   &	5.64-11  &   5.257-11  & 5.356-11   &  5.28-11     \\
   22   & 2p$^5$3d  &	$^3$D$^o_{3}$	  &   6.693-11  & 6.842-11   & 6.76-11    &  6.330-11  & 6.465-11   &	6.39-11  &   5.997-11  & 6.118-11   &  6.05-11     \\
   23   & 2p$^5$3d  &	$^3$D$^o_{1}$	  &   6.693-14  & 6.535-14   & 6.85-14    &  5.099-14  & 5.010-14   &	5.19-14  &   3.958-14  & 3.910-14   &  4.01-14     \\
   24   & 2p$^5$3d  &	$^3$F$^o_{2}$	  &   6.036-11  & 6.157-11   & 6.06-11    &  5.608-11  & 5.713-11   &	5.62-11  &   5.208-11  & 5.302-11   &  5.22-11     \\
   25   & 2p$^5$3d  &	$^3$D$^o_{2}$	  &   6.906-11  & 7.062-11   & 6.96-11    &  6.555-11  & 6.694-11   &	6.60-11  &   6.226-11  & 6.352-11   &  6.26-11     \\
   26   & 2p$^5$3d  &	$^1$F$^o_{3}$	  &   7.068-11  & 7.217-11   & 7.12-11    &  6.696-11  & 6.830-11   &	6.74-11  &   6.352-11  & 6.473-11   &  6.39-11     \\
   27   & 2p$^5$3d  &	$^1$P$^o_{1}$	  &   2.548-14  & 2.708-14   & 2.48-14    &  2.208-14  & 2.342-14   &	2.15-14  &   1.929-14  & 2.041-14   &  1.88-14     \\

\\  \hline            								                	 
\end{longtable}   								   					       
			      							   					       
%\vspace*{0.5 cm}													       

%}													       
														       
\begin{flushleft}													       
{\small
GRASP$^a$: Present calculations with the GRASP code \\
GRASP$^b$: Calculations of Jonsson et al. \cite{jon} with the GRASP2K code \\
GRASP$^c$: Calculations of Singh and Aggarwal \cite{sin} with the GRASP code \\
}															       
\end{flushleft} 

%\end{document}
\clearpage
\newpage

%%% Please start a new page by uncommenting the next

\TableExplanation

\bigskip
\renewcommand{\arraystretch}{1.0}

\section*{Table 1.\label{tbl1te} Energies (Ryd) for the lowest 139 levels of Cu~XX and their lifetimes ($\tau$, s).}% For the ground level the energy is absolute whereas for others  are comparative.}
\begin{tabular}{@{}p{1in}p{6in}@{}}
Index            & Level Index \\
Configuration    & The configuration to which the level belongs \\
Level             & The $LSJ$ designation of the level \\
GRASP          & Present energies from the GRASP code  with 64  configurations and 3948 level calculations \\
FAC               & Present energies from the FAC code  with 93~437 level calculations \\
$\tau$ (s)       & Lifetime of the level in s with the GRASP code \\

\end{tabular}
\label{tableII}

\bigskip
\renewcommand{\arraystretch}{1.0}

\section*{Table 2.\label{tbl2te} Energies (Ryd) for the lowest 139 levels of Zn~XXI and their lifetimes ($\tau$, s).}% For the ground level the energy is absolute whereas for others  are comparative.}
\begin{tabular}{@{}p{1in}p{6in}@{}}
Index            & Level Index \\
Configuration    & The configuration to which the level belongs \\
Level             & The $LSJ$ designation of the level \\
GRASP          & Present energies from the GRASP code  with 64  configurations and 3948 level calculations \\
FAC            & Present energies from the FAC code  with 93~437 level calculations \\
$\tau$ (s)       & Lifetime of the level in s \\

\end{tabular}
\label{tableII}

\bigskip
\renewcommand{\arraystretch}{1.0}

\section*{Table 3.\label{tbl3te} Energies (Ryd) for the lowest  139 levels of Ga~XXII and their lifetimes ($\tau$, s).}% For the ground level the energy is absolute whereas for others  are comparative.}
\begin{tabular}{@{}p{1in}p{6in}@{}}
Index            & Level Index \\
Configuration    & The configuration to which the level belongs \\
Level             & The $LSJ$ designation of the level \\
GRASP          & Present energies from the GRASP code  with 64  configurations and 3948 level calculations \\
FAC               & Present energies from the FAC code  with 38~089 level calculations \\
$\tau$ (s)       & Lifetime of the level in s with the GRASP code \\

\end{tabular}
\label{tableII}

\bigskip
\renewcommand{\arraystretch}{1.0}

%The following serves as an example on how to format a complicated
%explanation of table. The corresponding Table is however missing in this
%template.
%\end{document}

\bigskip
\section*{Table 4.\label{tbl4te}  Transition wavelengths ($\lambda_{ij}$ in $\rm \AA$), radiative rates (A$_{ji}$ in s$^{-1}$),
 oscillator strengths (f$_{ij}$, dimensionless), and line strengths (S, in atomic units) for electric dipole (E1), and 
A$_{ji}$ for electric quadrupole (E2), magnetic dipole (M1) and magnetic quadrupole (M2) transitions of Cu~XX.
 The ratio R(E1) of velocity and length forms of A-values for E1 transitions is listed in the last column.}
\begin{tabular}{@{}p{1in}p{6in}@{}}
$i$ and $j$         & The lower ($i$) and upper ($j$) levels of a transition as defined in Table 1.\\
$\lambda_{ij}$      & Transition wavelength (in ${\rm \AA}$) \\
A$^{E1}_{ji}$       & Radiative transition probability (in s$^{-1}$) for the E1 transitions \\
f$^{E1}_{ij}$       & Absorption oscillator strength (dimensionless) for the E1 transitions \\
S$^{E1}$            & Line strength in atomic unit (a.u.), 1 a.u. = 6.460$\times$10$^{-36}$ cm$^2$ esu$^2$ for the E1 transitions \\
A$^{E2}_{ji}$       & Radiative transition probability (in s$^{-1}$) for the E2 transitions \\
A$^{M1}_{ji}$       & Radiative transition probability (in s$^{-1}$) for the M1 transitions \\
A$^{M2}_{ji}$       & Radiative transition probability (in s$^{-1}$) for the M2 transitions \\
R(E1)                     & Ratio of velocity and length forms of A- (or f- and S-) values for the E1 transitions \\
$a{\pm}b$ &  $\equiv a\times{10^{{\pm}b}}$ \\
\end{tabular}
\label{ExplTable4}

\bigskip
\section*{Table 5.\label{tbl5te}  Transition wavelengths ($\lambda_{ij}$ in $\rm \AA$), radiative rates (A$_{ji}$ in s$^{-1}$),
 oscillator strengths (f$_{ij}$, dimensionless), and line strengths (S, in atomic units) for electric dipole (E1), and 
A$_{ji}$ for electric quadrupole (E2), magnetic dipole (M1) and magnetic quadrupole (M2) transitions of Zn~XXI.
The ratio R(E1) of velocity and length forms of A-values for E1 transitions is listed in the last column.}
\begin{tabular}{@{}p{1in}p{6in}@{}}
$i$ and $j$         & The lower ($i$) and upper ($j$) levels of a transition as defined in Table 2.\\
$\lambda_{ij}$      & Transition wavelength (in ${\rm \AA}$) \\
A$^{E1}_{ji}$       & Radiative transition probability (in s$^{-1}$) for the E1 transitions \\
f$^{E1}_{ij}$       & Absorption oscillator strength (dimensionless) for the E1 transitions \\
S$^{E1}$            & Line strength in atomic unit (a.u.), 1 a.u. = 6.460$\times$10$^{-36}$ cm$^2$ esu$^2$ for the E1 transitions \\
A$^{E2}_{ji}$       & Radiative transition probability (in s$^{-1}$) for the E2 transitions \\
A$^{M1}_{ji}$       & Radiative transition probability (in s$^{-1}$) for the M1 transitions \\
A$^{M2}_{ji}$       & Radiative transition probability (in s$^{-1}$) for the M2 transitions \\
R(E1)                     & Ratio of velocity and length forms of A- (or f- and S-) values for the E1 transitions \\
$a{\pm}b$ &  $\equiv a\times{10^{{\pm}b}}$ \\
\end{tabular}
\label{ExplTable5}

\bigskip
\section*{Table 6.\label{tbl6te}  Transition wavelengths ($\lambda_{ij}$ in $\rm \AA$), radiative rates (A$_{ji}$ in s$^{-1}$),
 oscillator strengths (f$_{ij}$, dimensionless), and line strengths (S, in atomic units) for electric dipole (E1), and 
A$_{ji}$ for electric quadrupole (E2), magnetic dipole (M1) and magnetic quadrupole (M2) transitions of Ga~XXII.
 The ratio R(E1) of velocity and length forms of A-values for E1 transitions is listed in the last column.}
\begin{tabular}{@{}p{1in}p{6in}@{}}
$i$ and $j$         & The lower ($i$) and upper ($j$) levels of a transition as defined in Table 3.\\
$\lambda_{ij}$      & Transition wavelength (in ${\rm \AA}$) \\
A$^{E1}_{ji}$       & Radiative transition probability (in s$^{-1}$) for the E1 transitions \\
f$^{E1}_{ij}$       & Absorption oscillator strength (dimensionless) for the E1 transitions \\
S$^{E1}$            & Line strength in atomic unit (a.u.), 1 a.u. = 6.460$\times$10$^{-36}$ cm$^2$ esu$^2$ for the E1 transitions \\
A$^{E2}_{ji}$       & Radiative transition probability (in s$^{-1}$) for the E2 transitions \\
A$^{M1}_{ji}$       & Radiative transition probability (in s$^{-1}$) for the M1 transitions \\
A$^{M2}_{ji}$       & Radiative transition probability (in s$^{-1}$) for the M2 transitions \\
R(E1)                     & Ratio of velocity and length forms of A- (or f- and S-) values for the E1 transitions \\
$a{\pm}b$ &  $\equiv a\times{10^{{\pm}b}}$ \\
\end{tabular}
\label{ExplTable6}

\end{document}